%% file: ciclops2011.tex
\documentclass{llncs}

\usepackage{url}
\usepackage{verbatim}

\usepackage{graphicx}
\usepackage{psfrag}
\usepackage{alltt}

\begin{document}
\bibliographystyle{splncs03} \title{Using Constraint Handling Rules to
  Provide Static Type Analysis for the Q Functional Language}

\author{J\'{a}nos Csorba \and Zsolt Zombori  \and P\'{e}ter Szeredi}
\institute{Department of Computer Science and Information Theory\\
Budapest University of Technology and Economics \\
\email{\{csorba,zombori,szeredi\}@cs.bme.hu}}

%Budapest, Magyar tud\'{o}sok k\"{o}r\'{u}tja 2. H-1117, Hungary \\
%E-mail: \{zombori,csorba,szeredi\}@cs.bme.hu
%}

\maketitle

\begin{abstract}
We describe an application of Prolog: a type checking tool for the Q
functional language. Q is a terse vector processing language, a descendant
of APL, which is getting more and more popular, especially in financial
applications. Q is a dynamically typed language, much like
Prolog. Extending Q with static typing improves both the readability of
programs and programmer productivity, as type errors are discovered by the
tool at compile time, rather than through debugging the program execution.

The type checker uses constraints that are handled by Prolog Constraint
Handling Rules. During the analysis, we determine the possible type
values for each program expression and detect inconsistencies. As most
built-in function names of Q are overloaded, i.e. their meaning
depends on the argument types, a quite complex system of constraints
had to be implemented.
\end{abstract}

\section{Introduction}
Our paper presents ongoing work on the type analysis tool \emph{qtchk}
for the Q vector processing language. The tool has been developed in a
collaborative project between Budapest University of Technology and
Economics and Morgan Stanley Business and Technology Centre,
Budapest. We described our first results in \cite{ICLP2011}. That
version provided \emph{type checking}: the programmer was expected to
provide type annotations (in the form of appropriate Q comments) and
our task was to verify the correctness of the annotations. In the
current version we move from type checking towards \emph{type
  inference}: we no longer require any type annotations (although we
allow them), but infer the possible types of all expressions from the
program code. Consequently, for any syntactically correct Q program
the analyser will detect type inconsistencies, as well as list the
possible types for consistent expressions.

In Section~\ref{sec:background} we briefly introduce the Q language
and provide an overview of our type analysis tool. For more details,
we refer the readers to \cite{ICLP2011}. Afterwards, in
Section~\ref{sec:csp} we present the constraint satisfaction problem
(CSP) and argue that type reasoning can be seen as a
CSP. Section~\ref{sec:type_for_q} introduces the type system of the Q
language. Section~\ref{sec:implementation} is devoted to implementing
type inference in the qtchk program. In Section~\ref{sec:evaluation}
we provide an evaluation of the tool developed and give an outline of
future work, while in Section~\ref{sec:related} we review some
approaches related to our work. Finally, Section~\ref{sec:conclusion}
concludes the paper.

\input{ciclops_background}

\input{ciclops_csp}
\input{ciclops_type_for_q}
\input{ciclops_implementation}
\input{ciclops_evaluation}
\input{ciclops_related}

\section{Conclusions}
\label{sec:conclusion}
We are in the process of developing a type inference tool for the Q
language as a Prolog application. We build on our previous experiences
with a type checker for the same language. For type inference, we make
no restriction as to how much type information is provided by the
user. We determine for each program expression the set of possible
types and indicate inconsistencies as well as clashes with programmer
provided type declarations. The type inference is constraint based,
using the Prolog CHR library. Using constraints enabled us to capture
the highly polymorphic nature of built-in functions due to
overloading. 

\section*{Acknowledgements}

We acknowledge the support of Morgan Stanley Business and Technology
Centre, Budapest in the development of the Q type checker system.  We are
especially grateful to Bal\'azs G.\ Horv\'ath and Ferenc Bodon for their
encouragement and help.

\bibliography{ICLP2011}
\end{document}

%% file: ciclops_background.tex
\section{Background}
\label{sec:background}

In this section we first present the Q programming
language. Afterwards, we provide an overview of the qtchk type
analyser tool. Most of the text in this section is taken directly from
\cite{ICLP2011} which describes the first version of qtchk. For more
details about the Q language and the architecture of our system, we
refer the reader to \cite{ICLP2011}.

\subsection{The Q Programming Language}
Q is a highly efficient vector processing functional language, which is
well suited to performing complex calculations quickly on large
volumes of data.  Consequently, numerous investment banks
(Morgan Stanley, Goldman Sachs, Deutsche Bank, Zurich Financial Group,
etc.) use this language for storing and analysing financial time
series~\cite{Customers}. The Q language \cite{QForMortals} first appeared
in $2003$ and is now (April $2011$) so popular, that it is ranked among the 
top $30$ programming languages by the TIOBE Programming Community~\cite{Tiobe}.

\paragraph{Types}
Q is a strongly typed, dynamically checked language. This means that
while each variable is associated with a well defined type, the type
of a variable is not declared explicitly, but stored along its value
during execution. The most important types are as follows:
\begin{itemize}
\item \textbf{Atomic types} in Q correspond to those in SQL with some
  additional date and time related types that facilitate time
  series calculations. Q has the following $16$ atomic types: \texttt{boolean},
  \texttt{byte}, \texttt{short}, \texttt{int}, \texttt{long},
  \texttt{real}, \texttt{float}, \texttt{char}, \texttt{symbol},
  \texttt{date}, \texttt{datetime}, \texttt{minute}, \texttt{second},
  \texttt{time}, \texttt{timespan}, \texttt{timestamp}.
\item \textbf{Lists} are built from Q expressions of arbitrary
  types.
\item \textbf{Dictionaries} are a generalisation of lists and provide
  the foundation for tables. A dictionary is a mapping that is given
  by exhaustively enumerating all domain-range pairs. For example,
  \texttt{(`a`b ! 1 2)} is a dictionary that maps symbols \texttt{a,b}
  to integers \texttt{1,2}, respectively.
\item \textbf{Tables} are lists of special dictionaries called
  \textbf{records}, that correspond to SQL records.
\end{itemize}

\paragraph{Main Language Constructs}
Q being a functional language, functions form the basis of the
language. A function is composed of an optional parameter list and a body
comprising a sequence of expressions to be evaluated. Function
application is the process of evaluating the sequence of expressions
obtained after substituting actual arguments for formal parameters. 

As an example, consider the expression
\begin{verbatim}
f: {[x] $[x>0;sqrt x;0]}
\end{verbatim}
which defines a function of a single argument $x$, returning $\sqrt{x}$, if
$x>0$, and 0 otherwise. Note that the formal parameter specification
\texttt{[x]} can be omitted from the above function, as Q assumes
\texttt{x}, \texttt{y} and \texttt{z} to be implicit formal parameters.

%% If a return value is specified, the function evaluates to its return
%% value, otherwise it has no return value. 

Input and return values of functions can
also be functions: for example, a special group of functions, called
\emph{adverbs} take functions and return a modified version of the
input. 
%% The whole Q program can be seen as a series of complex function evaluation steps.

Some built-in functions (dominantly mathematical functions) with one
or two arguments have a special behaviour called \emph{item-wise
  extension}.  Normally, the built-in functions take atomic arguments
and return an atomic result of some numerical calculation. However,
these functions extend to list arguments item-wise. If a unary
function is given a list argument, the result is the list of results
obtained by evaluating each argument element. A binary function
with an atom and a list argument evaluates the atom with each list
element. When both arguments are lists, the function operates
pair-wise on elements in corresponding positions. Item-wise extension
applies recursively in case of deeper lists, e.g.\ \texttt{((1;2);
  (3;4)) + (0.1; 0.2) = ((1.1;2.1); (3.2;4.2))}

While being a functional language, Q also has imperative features, such as
multiple assignment variables, loops, etc.

%% \paragraph{Flexibility}
%% Q is an extremely permissive language: for example, it is allowed to
%% divide by zero and built-in functions accept extreme types without
%% runtime error. This property of the language significantly increases
%% the chance of program errors that are very difficult to explore once
%% the program evaluation fails. Overcoming this difficulty by developing
%% debugging     tools for Q is likely to greatly enhance the usability of
%% the language.

\paragraph{Type restrictions in Q}
The program code environment can impose various kinds of restrictions
on types of expressions. In certain contexts, only one type is
allowed. For example, in the do-loop \texttt{do[n;x*:2]}, the first
argument specifies how many times \texttt{x} has to be multiplied by
\texttt{2} and it is required to be an integer. In other cases we
expect a polymorphic type. If, for example, function \texttt{f} takes
arbitrary functions for argument, then its argument has to be of type
\texttt{ A -> B} (a function taking an argument of type \texttt{A} and
returning a value of type \texttt{B}), where \texttt{A} and \texttt{B} are
arbitrary types.
In the most general case, there is a restriction involving the types of
several expressions. For instance, in the expression \texttt{x = y +
  z}, the type of \texttt{x} depends on those of \texttt{y} and
\texttt{z}. A type analyser for Q has to use a framework that allows
for formulating all type restrictions that can appear in the program.

\subsection{Overview of the Qtchk Type Analyser}
The type analysis implemented in qtchk can be divided into three
parts:
\begin{itemize}
\item Pass 1: lexical and syntactic analysis \\ The Q program is parsed
  into an abstract syntax tree structure.
\item Pass 2: post processing \\ Some further transformations make the
  abstract syntax tree easier to work with.
\item Pass 3: type checking proper \\ The types of all expressions are
  processed, type errors are detected.
\end{itemize}

\begin{figure}[htbp]
  \psfrag{A}{\large{Abs}}
  \psfrag{T}{\large{Tree}}
  \psfrag{Q}{\large{Q program}}
  \psfrag{Tc}{\large{Type comments}}
  \psfrag{L}{\large{\hspace*{-0.15em}\bf Lexical}}
  \psfrag{An}{\large{\hspace*{-0.35em}\bf Analyser}}
  \psfrag{S}{\large{\hspace*{-0.2em}\bf Syntactic}}
  \psfrag{P}{\large{\hspace*{-0.1em}\bf Post}}
  \psfrag{Pg}{\large{\hspace*{-0.3em}\bf Processing}}
  \psfrag{E}{\large{Errors}}
  \psfrag{Tp}{\large{\hspace*{-0.2em}\bf Type}}
  \psfrag{C}{\large{\hspace*{-0.9em}\bf Reasoning}}
  \psfrag{Ts}{\large{types}}
  \psfrag{BIF}{\large{Built-in Func}}

  \includegraphics[angle=0,width=12cm]{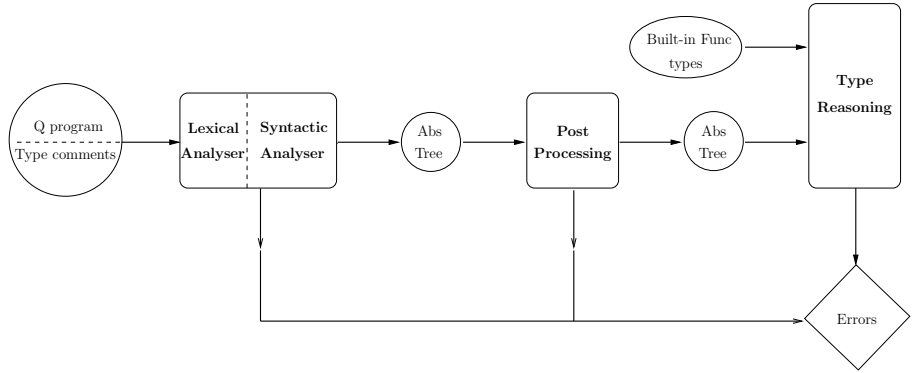}
\caption{Architecture of the type analyser}
\label{fig:architecture}
\end{figure}

The algorithm is illustrated in Figure~\ref{fig:architecture}. The
analyser receives the Q program along with the user provided type
declarations. The lexical analyser breaks the text into tokens. The
tokenizer recognises constants and hence their types are
revealed at this early stage. Afterwards, the syntactic analyser
parses the tokens into an abstract syntax tree representation of the Q
program. Parsing is followed by a post processing phase that
encompasses various small transformation tasks. 

In the post processing phase some context sensitive transformations are
carried out, such as filling in the omitted formal parameter parts in
function definitions, and finding, for each variable occurrence, the
declaration the given occurrence refers to.

Finally, in pass 3, the type analysis component traverses the abstract
syntax tree and imposes constraints on the types of the subexpressions
of the program. This phase builds on the user provided type
declarations and the types of built-in functions. The latter are
listed in a separate text file, that is parsed just like any Q
program. The predefined constraint handling rules trigger automatic
constraint reasoning, by the end of which the types (or the sets of
potential types) of all subexpressions are inferred.

Each phase of the type analyser detects and stores errors. At the end
of the analysis, the user is presented with a list of errors,
indicating the location and the kind of error. In case of type errors,
the analyser also gives some justification, in the form of conflicting
constraints.

In the rest of the paper, we describe improvement on the type analysis
component. The other parts of the system remain unchanged.

%% file: ciclops_csp.tex
\section{Type Inference as a Constraint Satisfaction Problem}
\label{sec:csp}
In this section we introduce the Constraint Satisfaction Problem
(CSP). Afterwards we present some general considerations on
translating type inference into a CSP. 

\subsection{Constraint Satisfaction Problem}
A constraint satisfaction problem (CSP) \cite{CSP} can be described with a
triple $(X, D, C)$, where 
\begin{itemize}
\item{} $X = \{x_1,\dots,x_n\}$ is a series of variables,
\item{} $D = \{D_1,\dots,D_n\}$ is a series of finite sets called
  domains,
\item{} variable $x_i$ can only take values from domain $D_i$,
\item{} $C = \{c_1,\dots,c_k\}$ is a series of constraints, i.e., atomic
relations whose arguments are variables from $X$.
\end{itemize}
A solution to a CSP is an assignment to each $x_i \in X$ a domain
element $v_i \in D_i$, such that all constraints $c \in C$ are
satisfied.

A value $d_i$ of a variable $x_i$ of a constraint $c$ is
\emph{superfluous} in case there is no assignment to the rest of the
variables of $c$ along with $x_i = d_i$ that satisfies constraint
$c$. Removing superfluous values from the corresponding domains yields
an equivalent CSP.

There are two mechanisms that lead to a solution of a CSP. First,
constraints constantly monitor the domains of their variables and
remove superfluous values. Second, in case constraints fail to reduce
some domain to a single value, we apply labeling: we choose a variable
$x_i$ and split its domain into two (or more) parts, creating a choice
point where each branch corresponds to a reduced domain. Through a
backtracking search we explore the branches. During labeling,
constraints can wake up as the domains of their variables change and
can further eliminate superfluous values. In case a domain becomes
empty, we roll back to the last choice point. By the end of labeling,
either we find a single value for each variable such that all
constraints are satisfied, or else we conclude that the CSP is
unsatisfiable.

\subsection{Type Inference and the Constraint Satisfaction Problem}
In this subsection we overview the requirements to transform type
reasoning into a CSP.

We start from a program code that can be seen as a complex expression
built out of simpler expressions. Our aim is to assign a type to each
expression appearing in the program in a coherent manner. The types of
some expressions are known immediately (atomic expressions, built-in
function symbols), while other types might be provided by the user
(through a type declaration). Besides, the program syntax imposes
restrictions that can be interpreted as constraints between the types
of certain expressions. A coherent type assignment respects all user
declarations and all constraints.

To each expression we assign a variable. In case the set $T$ of all
possible types is finite, we set the domain $D_i$ of variable $x_i$ to
$T$. If however, there are infinitely many types, we use type
expressions that represent sets of types. For example, the infinite
set of all homogeneous lists might be represented with the single 
polymorphic type expression \texttt{list(X)}. This opens the possibility 
to finitely represent infinite sets. For this, we have to design a type 
language such that for any set of types that is relevant for the 
programming language at hand we can provide a finite representation using 
type expressions. Let $T_k$ denote the set of types represented by type
expression $k$. For each variable $x_i$ we maintain a list $L_i$ of
type expressions, and we set the domain of the variable $D_i$ to
$\bigcup_{l \in L_i}T_l$.

For an expression with known type, we immediately restrict the domain
to the given value. Other restrictions appear as constraints that
monitor the domains of their variables and eliminate superfluous
values. Since a type expression stands in general for a set of types
and not for an individual type, narrowing the domain does not always
remove a type expression, but might also involve replacing it with
some other, depending on the particular constraint.

Even if we manage to finitely represent infinite sets of types, we
might still run into difficulty during labeling. By repeatedly
splitting the domain of a variable, we cannot guarantee that the
domain eventually turns into a singleton. Hence, instead of splitting
the domain $D_i$, we split the list of type expressions $L_i$. Once
this list becomes a singleton, we terminate labeling. Consequently, we
obtain (potentially infinite) sets of types for our expressions. This
is no problem, however, if the type expressions are chosen
carefully. Our first aim with labelling is not to obtain a unique type
for each expression, but to enable the constraints to wake up and
eliminate superfluous values. If the type expressions are ``fine
grained'' enough, such that constraints can exit once the types of
their arguments are all represented with a single type expression,
then there will be no constraints left by the end of labeling and we
can return the set of types corresponding to the type expression.

In conclusion, we formulate the following requirements towards a type
language to be used for type inference:
\begin{enumerate}
\item Each set of types that can be associated with an expression of
  the given programming language should be representable with a finite
  list of type expressions.
\item For each constraint $c$, if each of its variables $x_i$ is
  associated with a singleton list $L_i$ of type expressions, then $c$
  can exit.
\end{enumerate}
Given such a type language, we can treat the task of type inference as
a CSP. The only differences are that 1) instead of a set of types, we
maintain a finite set of type expressions to represent the domain of a
variable and 2) constraints not only remove, but sometimes replace
type expressions when eliminating superfluous values.

%% file: ciclops_type_for_q.tex
\section{Type Inference for the Q Language}
\label{sec:type_for_q}
After the general remarks in the previous section, we now examine the
Q specific aspects of type inference in the context of CSP.
\subsection{Type expressions}

We describe the type language developed for Q. A significant
improvement from the first type language presented in \cite{ICLP2011}
is that we allow polymorphic type expressions, i.e., any part of a
complex type expression can be replaced with a variable. Expressions
are built from atomic types and variables using type constructors. The
abstract syntax of the type language -- which is at the same time the
Prolog representation of types -- is as follows:

\begin{alltt}
    \textit{TypeExpr} =
          {\bf\rm{AtomicTypes}}
        | \textit{TypeVariable}
        | list(\textit{TypeExpr})
        | hlist
        | tuple([\textit{TypeExpr}, \ldots ,\textit{TypeExpr}])
        | stuple([\textit{Name}, \ldots ,\textit{Name}])
        | dict(\textit{TypeExpr}, \textit{TypeExpr})
        | func(\textit{TypeExpr}, \textit{TypeExpr})
\end{alltt}
where \textit{TypeVariable} is a Prolog variable. Due to the presence
of variables, a type expression represents a (possibly infinite) set
of types, which we take to be the meaning of the expression. In the
following, we list the meaning of all type expressions:

\begin{description}
\item[{\bf\rm{AtomicTypes}}]: This is shorthand for the 16 atomic types
  of Q.
\item[TypeVariable] The set of all expressions, with the restriction
  that the same variables need to stand for the same type expression.
\item[list(\textit{TE})] The set of all lists whose elements are all
  from the set represented by \textit{TE}.
\item[hlist] The set of all lists.
\item[tuple({[}\textit{TE$_1$}, \ldots , \textit{TE$_k$}{]})] The set of
  all lists of length $k$, such that the $i^{th}$ element is from the
  set represented by \textit{TE$_i$}.
\item[stuple({[}\textit{name$_1$}, \ldots , \textit{name$_k$}{]})] A
  singleton set consisting of the $k$ long symbol list whose $i^{th}$
  element is \textit{name$_i$}.
\item[dict(\textit{TE$_1$}, \textit{TE$_2$})] The set of all
  dictionaries, such that the domain and range are from the sets
  represented by \textit{TE$_1$} and \textit{TE$_2$},
  respectively. Domains and ranges are represented as a sequence of
  possible values, i.e., for example, the dictionary \texttt{(1.2 1.3 ! 1
    2)} has type \texttt{dict(tuple([float, float]), tuple([int,
      int]))}.
\item[func(\textit{TE$_1$}, \textit{TE$_2$})] The set of all
  functions, such that the domain and range are from the sets
  represented by \textit{TE$_1$} and \textit{TE$_2$}, respectively.
\end{description}

\paragraph{Mapping Type Inference to CSP}
For each Q expression we maintain a set of possible types, its
domain. As described in Section~\ref{sec:csp}, it is not the domain of
the Q expression that we keep track of, but a list of type
expressions. The domain can be obtained by taking the union of the
sets represented by the type expressions in the list. It is this list
that we try to narrow down as much as possible during constraint
reasoning.

\paragraph{Non-Overlapping Type Expressions}
While some type expressions correspond directly to Q language
constructs (such as \texttt{list}, \texttt{dict} or \texttt{func}),
others were ``discovered'' in the process of trying to describe Q
expressions. Such are the \texttt{tuple} and \texttt{stuple} type
expressions. Some built-in functions require list arguments with fixed
length. These lists might also have to be non-homogeneous, with well
specified type for each list member. To be able to describe the type of
such functions (and that of their argument), we introduced the
\texttt{tuple} type. Using the \texttt{tuple} type, we can for example easily
describe a function that takes a list consisting of an integer and a
symbol and returns another list consisting of two integers and a
float: \texttt{func(tuple([int,symbol]),tuple([int,int,float]))}.

An \texttt{stuple} is a degenerate \texttt{tuple} as it represents a
singleton set. This expression is necessary for manipulating tables:
if for instance we want the type checker to verify that a given record
can be inserted into a given table, then we have to know if the
record and the table have the same column names. A record is a
dictionary that maps column names to values. By using the
\texttt{stuple} type, we can represent the domain type of the
dictionary in such a way that contains the names of all
columns. Hence, instead of treating the dictionary \texttt{`name`age !
  (`jim;12)} as a
\texttt{dict(tuple([symbol,symbol]),tuple([symbol,int]))}, we represent
its type as \texttt{dict(stuple([name,age]),tuple([symbol,int]))}.
                
Introducing these types causes some difficulties, because type
expressions using different constructors are not necessarily
disjoint. As we have seen, a tuple is a special list, an stuple is a
special tuple. In the course of type inference, as constraints narrow
down the domains of expressions, we cannot use unification such simple to 
obtain a narrower set of types. As an example, consider the next two type
expressions: $T_1$ : \texttt{list(X)}, $T_2$ : \texttt{tuple([int, Y])}. 
Unification of these two terms leads to failure, however, the two expressions 
are not disjoint. If expression $E$ has to satisfy $T_1$ and $T_2$ 
simultaneously, the domain of $E$ has to be narrowed to 
\texttt{tuple([int, int])}, with the substitutions \texttt{X=int} and 
\texttt{Y=int}. This result can be obtained, for example, by unification of 
\texttt{tuple([X, X])} and \texttt{tuple([int, Y])}.

%\subsection{Extending Functions to List Arguments}

\subsection{Type Declarations}

When we first developed a type checker tool for Q \cite{ICLP_full},
the user was required to provide every variable and user-defined
function with a ground type description. Since then, we lifted both
parts of the restriction: 1) type declarations are not obligatory
and 2) we allow polymorphic type expressions using variables. However,
the user can still opt to provide a type annotation for an arbitrary
expression. Such annotations appear as Q comments and hence do not
interfere with the Q compiler. A type declaration gets attached to the
smallest expression that it follows immediately. For example, in the
code \texttt{x + y //\$: int} variable \texttt{y} is declared to be an
integer.

Type declarations can be of two kinds, having slightly different
semantics: \emph{imperative} (believe me that the type of expression E is T) or
\emph{interrogative} (I think the type of E is T, but please do
check). To understand the difference, suppose the value of \texttt{x}
is loaded from a file. This means that both the value and the type is
determined in runtime and the type checker will treat the type of
\texttt{x} as \texttt{any}. If the user gives an imperative type
declaration that \texttt{x} is a list of integers, then the type
analyser will believe this and treat \texttt{x} as a list of
integers. If, however, the type declaration is interrogative, then the
type analyser will issue a warning, because there is no guarantee that
\texttt{x} will indeed be a list of integers (it can be
anything). Interrogative declarations are used to check that a piece
of code works the way the programmer intended. Imperative declarations
provide extra information for the type analyser. 

Different comment tags have to be used for introducing the two kinds
of declarations. We give an example for each:
\begin{verbatim}
f //$: int -> boolean      interrogative
g //!: int -> int          imperative
\end{verbatim}

%% file: ciclops_implementation.tex
\section{Implementing Type Inference in the Qtchk program}
\label{sec:implementation}

The type checking tool has been implemented in SICStus Prolog 4.1
\cite{SICStus}. As it is described in more detail in \cite{ICLP_full},
we use a parser to build an abstract syntax tree from the Q program,
which is the input of the type analyser component. The output is the
list of expressions that contain type errors. During execution, we try
to assign a type to each program expression. Each expression is
represented by a node in the abstract syntax tree. In order to be able
to comfortably refer to various expressions, we extend each abstract
syntax tree node with a globally unique identifier. We use these
identifiers instead of variables in the CSP, i.e., each identifier
gets associated with a domain of type expressions. Besides, the
arguments of type constraints that we will formulate are
identifiers. Identifiers have to be provided to constraints in order
to be able to provide error messages pointing to a specific location,
and in the presence of identifiers, introducing new CSP variables is
unnecessary.

\subsection{Constraints}
                
Constraints are handled using the Prolog CHR \cite{CHR_library}
library.  As we have said earlier, node identifiers play the role of
CSP variables and our aim is to find a type for each identifier. We
represent domains using the CHR constraint \texttt{dom(ID, T)}, which
associates identifier \texttt{ID} with the list of type expressions
\texttt{T}. The constraint means that the type of the expression
identified by \texttt{ID} belongs to the set represented by
\texttt{T}. In case an expression is unconstrained, we do not add the
\texttt{dom/2} constraint.  This reduces the number of constraints.

In contrast to the earlier type checker tool, the order in which
constraints are added is irrelevant. It does not matter if some type
declaration is missing or if we first constrain an expression and then
constrain its subexpressions, or the other way around. Each constraint
is bound to do some narrowing in the domains of the identifiers it is
attached to, whether it comes later or earlier. Constraints that can
be used for type inference can originate from the following sources in
a Q program:

\begin{description}
\item[Imperative type declarations] If the user gives an imperative
  type declaration, then the type analyser will accept this
  unconditionally. This means that we add a \texttt{dom/2} constraint
  on the given expression.
\item[Built-in functions] For every built-in function, there is a
  well-defined relation between the types of its arguments and the
  type of the result. These relations are expressed by adequate CHR
  constraints. For each built-in function we provide manually a number
  of constraint handling rules to describe how the constraint is
  supposed to narrow domains. For example, we use the constraint
  \texttt{sum\_c} to capture the relation between the arguments of the
  built-in function `+`. So, if we see an expression \texttt{c:a+b},
  we add the constraint \texttt{sum\_c($id_a$,$id_b$,$id_c$)}.
\item[Atomic expressions] The types of atomic expressions are revealed
  already by the parser, so for example, \texttt{2.2f} is immediately
  known to be a float.
\item[Variables] Local variables are made globally unique by the
  parser. This means, that variables with same name are equal, hence
  their types are also equal. We ensure this by equating their
  corresponding domains.
\item[Program syntax] Most syntactic constructs impose some
  constraints on the types of their constituent constructs. For
  example, the first argument of an \texttt{if-then-else} construct
  must be a boolean value. Another example is function application:
  it has a subexpression with type \textit{func(a,b)}, another
  subexpression with type \textit{a} and the whole expression is of
  type \textit{b}.
\end{description}

\subsection{Item-wise List Extension of Built-in Functions}
Capturing the item-wise extension of built-in functions requires
further considerations. When we see the expression $c:a + b$, then
either $a$ and $b$ have atomic types and the $'sum'$ relation applies
to them, or at least one of them is a list and the relation applies to
the list elements. One way to capture this is to make the constraints
clever enough, i.e., simply add the constraint
\texttt{sum\_c(id(a),id(b),id(c))} and provide the adequate rules for
the \texttt{sum\_c} constraint. The disadvantage of this approach is
that the rules describing the list extension behaviour have to be
repeated for each and every built-in function, which is not
productive. Instead, we introduced a metaconstraint
\texttt{listextension/3}.  Let $f$ be a binary built-in function,
which extends item-wise to lists in both arguments and which imposes
constraint \texttt{c} on its atomic arguments and result. As we
traverse the abstract syntax tree, suppose we meet $f$ with arguments
identified by $ID1$, $ID2$ and result identified by $ID3$. We cannot
add \texttt{c(ID1,ID2,ID3)} to the constraint store until we know for
sure that the the arguments are all of atomic type. Instead, we use
the metaconstraint \texttt{listextension(Dir, Args, Cons)}, where
\texttt{Dir} specifies which arguments can be extended item-wise to
lists, \texttt{Args} is the list of arguments on which the list of
constraints \texttt{Cons} will eventually have to be
formulated. Hence, in our example, we add the constraint
\texttt{listextension(both,[ID1,ID2,ID3],[c])}. If we somehow infer
that the input arguments are atomic, then we simply add the constraint
\texttt{c(ID1,ID2,ID3)} and the metaconstraint can exit. If, on the
other hand some argument turns out to be a list, we replace the
metaconstraint with another one. For example, if we know that the type
of $ID1, ID2$ are \texttt{list(A)} and \texttt{list(B)}, respectively, 
then the type of $ID3$ must be a list as well and we replace our 
listextension constraint with the following two constraints:
\texttt{listextension(both,[A,B,C],[c])}, \texttt{dom(ID3,[list(C)])}.

Using the \texttt{listextension/3} metaconstraint provides a recursive
solution that can handle lists of arbitrary depth and that treats all
extendable functions in a uniform manner.

We illustrate list extension with the simple Q program:
\texttt{c:c+1}. The corresponding abstract syntax tree is as follows:
\begin{verbatim}
                            assign
                             id(1)
                          /        \
                       var(c)      app
                       id(2)      id(3)
                                /       \
                              var(+)    list
                              id(4)     id(5)
                                       /     \
                                     var(c)  int
                                     id(6)   id(7)
\end{verbatim}

Table~\ref{tab:imp_example} summarises the added constraints. As we
reach the \texttt{assign} node, we know that the type of the left side
(\texttt{id(2)}) is the same as the type of the right side
(\texttt{id(3)}) which also equals the type of the whole assignment
(\texttt{id(1)}). Later, when we find the second occurrence of
variable \texttt{c}, we know that its type must equal with the type of
the first occurrence of \texttt{c}. Once we reach the \texttt{+}
function, we add the \texttt{listextension/3} metaconstraint.  The
number \texttt{1} is immediately recognised as an integer. Let us
suppose that variable \texttt{c} later turns out to be a list, i.e.,
of type \texttt{list(X)}. Hence the result of the sum (\texttt{id(3)})
must also be a list (\texttt{list(Z)}), and the metaconstraint has to be
formulated on the list members. Finally, suppose \texttt{X} turns out
to be a float. Then, \texttt{listextension} can be replaced with the
\texttt{sum\_c} constraint, which will now have atomic arguments and
which will exit after setting the domain of \texttt{Z} to float.

\begin{table}[htb]
        \caption{Constraints related to the expression \texttt{c:c+1}.}
        \label{tab:imp_example}
        \center
        \begin{tabular}{|c|c|}
        \hline
        \emph{Reason} & \emph{Constraints} \\
        \hline
        node \texttt{assign} & \texttt{eq(id(2),id(3)), eq(id(1),id(3))} \\
        node \texttt{app} & \texttt{dom(id(4), [func(id(6),id(7),id(3))])} \\
        variable $c$ &  \texttt{eq(id(2),id(6))} \\
        function $+$ & \texttt{listextension(both,[id(6),id(7),id(3)],[sum\_c])} \\
        constant $1$ & \texttt{dom(id(7),[int])} \\
        \texttt{dom(id(6),[list(X)])} &
        \texttt{listextension(both,[X,id(7),Z],[sum\_c])} \\
        & \texttt{dom(id(3),[list(Z)])} \\
        \texttt{dom(X,[float])} & \texttt{sum\_c(X,id(7),Z)} \\
        \texttt{sum\_c(X,id(7),Z)} & \texttt{dom(Z,[float])} \\
        \hline
        \end{tabular}
\end{table} 
 
\subsection{Constraint Interaction}
The CHR constraint \texttt{dom/2} is used to represent the domains of
constraint-variables (that are represented in our solution with
identifiers). The domain is a list of type expressions, that are
Prolog structures including variables. Other constraints interact
directly with the \texttt{dom/2} constraints. Different constraints
work together through making changes in the \texttt{dom/2}
constraints. Changing the domain always results in a domain, where the
new list of Prolog expressions represents a narrower set of types than
the original one. This does not necessarily mean that the size of the
domain list is reduced, as is the case when the type of an expression
is refined from \texttt{list(X)} to \texttt{list(int)} or even to
\texttt{tuple(int,int)}.

Constraints do not exclusively narrow variable domains. In some
situations it is necessary for constraints to invoke other constrains.
For example, if one argument of the \texttt{sum\_c} constraint turns
out to be an integer, then the analyser infers that the type of the
other argument and the result must equal.
                
\subsection{Labeling}

The inferred type of a Q expression is described in a list of type
expressions using the \texttt{dom/2} constraint. Once all constraints
have been added, labeling might be necessary. This subsection is
somewhat speculative, because we have not yet implemented labeling in
our system.

The constraints might remain suspended, however, they are guaranteed
to exit once the domain lists of their arguments become singleton. We
apply labeling to ensure that the suspended constraints are not
inconsistent. Hence, our aim with labeling is not to assign an unique
type to each expression, rather to split the domains until all
constraints exit. During labeling, we split the domain lists and stop
once all lists become singletons (note that in this case the real
domain associated with the expression is a set, which can as well be
infinite). This much labeling ensures that no constraints remain. For
example, if the domain of an expression is \texttt{list(X)}, and there
aren't any constraints on \texttt{X}, then we do not need further
labeling on this domain.

Individual constraints control the labeling. Once labeling starts, the
constraints examine the domains of their arguments and split them as
much as they need to in order to be able to exit. Hence, each
constraint is equipped with adequate rules for labeling. This solution
ensures on one hand that labeling goes on until active constraints
remain, and on the other that labeling stops as soon as there are no
active constraints left.

After labeling, either we find that the suspended constraints are 
inconsistent and issue an error message, or we collect the possible sets 
of types to each expressions. We show these sets to the user.
\subsection{Detection of Type Errors}

Type errors can be of two different kinds. On the one hand, the Q
program itself might contain an error (by using, for example, a float
where integer is expected). The type analyser detects these errors
when some domain reduces to the empty set. On the other hand, there
might be a mismatch between the inferred type of an expression and the
type provided by the user through an interrogative type
declaration. We examine both cases.

\subsubsection{Inferred Type Errors}
After having added all constraints, an empty domain of an expression
indicates a type error. However, type errors propagate upwards in the
abstract syntax tree, since if an expression is inconsistent, then so
is any superexpression as well. To avoid overburdening the user with
unnecessary details, we would like to point to the narrowest
expression that contains the error, so we will not list every
expressions with empty domain. The narrowest expression is the one
furthest down the abstract syntax tree. Hence, we show the expressions
that are inconsistent (their domains are empty) and whose children are
all consistent (their domains are not empty).

We note that there exists one abstract node for which this solution is
not fully satisfactory, the \texttt{assign} node. The assign node is the 
abstract form of the assignment expressions. The problem is that
here the type error not only propagates upwards, but sideways as well,
toward a sibling. In the expression \texttt{a:b}, if the domain of
\texttt{b} reduces to the empty set, then so will the domain of
\texttt{a}, even though none is the child of the other. In this
situation it is not necessary to indicate type error for \texttt{a}.
However, we cannot distinguish this case from the one in which
\texttt{a} contains a real type error (regardless of \texttt{b}), so
we decided to allow this unnecessary error message.  As an example,
consider the expression \texttt{l[2]: a+b}. Let us suppose that
variable \texttt{a} and variable \texttt{b} are lists with different
length, which implies a type error at the right side of the
\texttt{assign} node. If the variable \texttt{l} is a list, then it is
not necessary to indicate type error on the left side of the
assignment. However, if variable \texttt{l} turns out to be a
function, we have a real type error in the subexpression
\texttt{l[2]}. Note that in both cases we obtain the same set of nodes
with empty domains.

Some type errors might remain hidden until we do labeling. In this
case, labeling will fail and we will know there is some error, but we
will not know its location. We are still working to provide a useful
error message for the user in these situations.

\subsubsection{Interrogative type declarations}
Interrogative declarations are used to check that a piece of code works the 
way the programmer intended. The user can ask for any expression if
its type is guaranteed to be the one he expects.

Let $T_1$ be the set of types inferred for expression $E$ and $T_2$
the set of types provided by an interrogative type declaration. The type
analyser has to distinguish the following cases:
\begin{itemize}
\item
$T_1$ and $T_2$ are disjoint. The analyser has to issue a type error.
\item
$T_1$ and $T_2$ intersect, but there exist some element of $T_1$ which
  is not element of $T_2$. In this case, the program might run
  correctly, but there is no guarantee. We indicate this by issuing a
  type warning.
\item
$T_1$ is subset of $T_2$. This means that the program satisfies the
  expectations of the programmer and no error message is necessary.
\end{itemize}

Sets $T_1$ and $T_2$ are represented with lists of type expressions
that can be polymorphic and that might further be constrained by all
sorts of constraints. In this generic scenario, it is very difficult to
determine the exact relation of the two sets and we have yet to come
up with a satisfactory solution.

%% file: ciclops_evaluation.tex
\section{Evaluation and Future Work}
\label{sec:evaluation}

A static type checking tool, described in \cite{ICLP2011}, has been
developed in Prolog in about 6 months by the three authors of this
paper. While that version is under evaluation on real-life Q programs
at Morgan Stanley Business and Technology Centre, we have started
developing a new mechanism for type analysis, that allows us to move
from type checking towards type inference. In the first version we put
the emphasis on completeness: our analyser could determine the unique
type of each program expression. To achieve this, the user was
required to provide type annotations for all user defined functions
and all variables. In the new version described currently, we ease the
burden of the programmer and let him declare as many types as he
pleases. As a result, we cannot always determine the exact type of all
expressions. The program infers as much as can be inferred: if there
is a certain type error, we indicate the error; if there is a clash
between the inferred type and the declared type, we again issue an
error; if there is an expression whose type cannot be determined, we
provide the set of types from which it takes value. Our system can be
used for checking type declarations as well as for zero knowledge type
inference, depending on the information provided by the programmer.

Using CHR for type reasoning turned out to be very convenient. We can
represent the possible types for an expression using the
\texttt{dom/2} constraint. CHR also allows us to change the type of an
expression during reasoning -- for instance, what first was a
\texttt{list(int)} later turns out to be a \texttt{tuple(int,int,int)}
-- which would be very difficult to achieve in a unification based
inference mechanism.

We also extended the type language by allowing polymorphic type
expressions, which allows the programmer to describe much more
types. We further plan to allow stating constraints on the variables
of the type declaration, however, this task is yet to be explored.

There are still lots of open questions. Maybe most importantly, it is
not clear what to do once the types of some expressions remain
ambiguous, with suspended constraints attached to them. We can wake up
the constraints with some sort of labeling, however, it is not clear
how to interpret the type of an expression obtained after labeling,
and how to compare it with the declared type of the expression.

A lot of constraints related to various built-in functions still need
to be implemented, which promises to be a tedious, but rather
straightforward work.

Working with ambiguous types can potentially result in lots of
suspended constraints and large search space to be explored during
labeling. Our system is not yet in the test phase, so it is still an
open question what sort of performance difficulties we will have to
cope with.

%% file: ciclops_related.tex
\section{Related Work}
\label{sec:related}

Several dynamically typed languages have been extended with a type
system allowing for static type checking or type
inference. \cite{MycroftO84} describe a polymorphic type system for
Prolog. \cite{Marlow&Wadler} present a type system for Erlang, which
is similar to Q in that they are both dynamically typed functional
languages. Several of the shortcomings of this system were addressed
in \cite{LindahlS06}. The tool presented in this work differs from
ours in its motivation. It requires no alteration of the code (no type
annotations) and infers function types from their usage. Instead of
well-typing, it provides success typing: it aims to discover provable
type errors. We, on the other hand, search for potential
errors. \cite{Demoen98} report on using constraints in type checking
and inference for Prolog. They transform the input logic program with
type annotations into another logic program over types, whose
execution performs the type checking. They give an elegant solution to
the problem of handling infinite variable domains by not explicitly
representing the domain on unconstrained variables. We borrowed this
idea and introduced type expressions to finitely represent infinite
domains. \cite{Sulzmann08} describe a generic type inference system
for a generalisation of the Hindley-Milner approach using constraints,
and also report on an implementation using Constraint Handling Rules.
% http://ww2.cs.mu.oz.au/~pjs/papers/jfp2007.pdf